\documentclass[aip,amsmath,amssymb,amsfonts,reprint]{revtex4-1}

\usepackage{graphicx}
\usepackage{dcolumn}
\usepackage{bm}
\usepackage[mathlines]{lineno}

\usepackage[utf8]{inputenc}
\usepackage[T1]{fontenc}
\usepackage{mathptmx}
\usepackage{etoolbox}
\usepackage{pifont}
\usepackage{url}

\DeclareMathAlphabet{\mathpzc}{OT1}{pzc}{m}{it}

\makeatletter
\def\@email#1#2{%
 \endgroup
 \patchcmd{\titleblock@produce}
  {\frontmatter@RRAPformat}
  {\frontmatter@RRAPformat{\produce@RRAP{*#1\href{mailto:#2}{#2}}}\frontmatter@RRAPformat}
  {}{}
}%
\makeatother

\makeatletter

\newcommand{\fmarkii}{\ensuremath{\dagger}}
\newcommand{\fmarkiii}{\ensuremath{\ddagger}}
\newcommand{\fmarkiv}{\ensuremath{\mathsection}}
\newcommand{\fmarkv}{\ensuremath{\mathparagraph}}
\newcommand{\fmarkvi}{\ensuremath{\|}}
\newcommand{\fmarkvii}{**}
\newcommand{\fmarkviii}{\ensuremath{\dagger\dagger}}
\newcommand{\fmarkix}{\ensuremath{\ddagger\ddagger}}
\def\@fnsymbol#1{{\ifcase#1\or \fmarkii\or \fmarkiii\or \fmarkiv\or \fmarkv\or \fmarkvi\or \fmarkvii\or \fmarkviii\or \fmarkix \else\@ctrerr\fi}}
\makeatother
\newcommand{\STO}{SrTiO$_3$ }

\begin{document}

\preprint{}

\title[Stoichiometric growth of SrTiO$_3$ films via Bayesian optimization with adaptive prior mean]{Stoichiometric growth of SrTiO$_3$ films \\via Bayesian optimization with adaptive prior mean}

\author{Yuki K. Wakabayashi}
 \thanks{These authors equally contributed to this work.\newline {\scriptsize {\bf Authors to whom correspondence should be addressed}: \texttt{yuuki.wakabayashi.we@hco.ntt.co.jp}, \texttt{takuma.otsuka.uf@hco.ntt.co.jp}}}
 \affiliation{
 NTT Basic Research Laboratories, NTT Corporation, Atsugi, Kanagawa 243-0198, Japan
 }

\author{Takuma Otsuka}%
 \thanks{These authors equally contributed to this work.\newline {\scriptsize {\bf Authors to whom correspondence should be addressed}: \texttt{yuuki.wakabayashi.we@hco.ntt.co.jp}, \texttt{takuma.otsuka.uf@hco.ntt.co.jp}}}
\affiliation{ 
NTT Communication Science Laboratories, NTT Corporation, Seika-cho, Kyoto 619-0237, Japan
}%

\author{Yoshiharu Krockenberger}
 \affiliation{
 NTT Basic Research Laboratories, NTT Corporation, Atsugi, Kanagawa 243-0198, Japan
 }
\author{Hiroshi Sawada}
\affiliation{
NTT Communication Science Laboratories, NTT Corporation, Seika-cho, Kyoto 619-0237, Japan
}
\author{Yoshitaka Taniyasu}
\author{Hideki Yamamoto}
 \affiliation{
 NTT Basic Research Laboratories, NTT Corporation, Atsugi, Kanagawa 243-0198, Japan
 }


\begin{abstract}
Perovskite insulator \STO is expected to be applied to the next generation of electronic and photonic devices as high-$k$ capacitors and photocatalysts. However, reproducible growth of highly insulating stoichiometric \STO films remains challenging due to the difficulty of the precise stoichiometry control in perovskite oxide films. 
Here, to grow stoichiometric \STO thin films by fine-tuning multiple growth conditions, we developed a new Bayesian optimization (BO)-based machine learning method that encourages the exploration of the search space by varying the prior mean to get out of suboptimal growth condition parameters. 
Using simulated data, we demonstrate the efficacy of the new BO method, which reproducibly reaches the global best conditions. 
With the BO method implemented in machine-learning-assisted molecular beam epitaxy (ML-MBE), highly insulating stoichiometric \STO film with no absorption in the band gap was developed in only 44 MBE growth runs. 
The proposed algorithm provides an efficient experimental design platform that is not as dependent on the experience of individual researchers and will accelerate not only oxide electronics but also various material syntheses.
\end{abstract}

\maketitle


\section{\label{sec:intro}Introduction}
Perovskite insulator \STO (STO) (cubic structure with the lattice constant of 3.905~\AA), having a band gap of 3.2~eV, is one of the most promising materials for oxide electronics.\cite{01shi2020,02pai2018,03fujimoto1985,04sakuma1998}
It is expected to be applied to high-$k$ capacitors\cite{04sakuma1998,05galt1993} and photocatalyst\cite{06mav1976,07konta2004,08ng2010} owing to its high dielectric constant of 100--200 at room temperature,\cite{04sakuma1998,09lee2008} 
chemical stability,\cite{01shi2020} almost 100\% quantum efficiency of photocatalytic water splitting under ultraviolet light (UV),\cite{06mav1976,10takata2020} and compatibility with other perovskite oxides.\cite{02pai2018,11ohtomo2004,12bowen2008,13wakabayashi2021a,14takada2021,15wakabayashi2021b,add_takada2022}
In addition, when it is doped by cation substitution, adding oxygen vacancies or cation vacancies, many interesting physical properties or phenomena emerge, such as superconducting states,\cite{16koonce1967,17bert2011,18li2011,19ahadi2019} ferroelectricity,\cite{20jang2010} high mobility carriers,\cite{21son2010,22matsubara2016} and blue light emission.\cite{23kan2005} 
However, mid-gap states originating from off-stoichiometry defects, such as oxygen and cation vacancies, are known to cause leakage current in STO capacitors,\cite{24popescu2014,25baek2020} 
and also cause mid-gap absorption that may decrease the photocatalytic activity of STO.\cite{26chen2018,27kumar2020} 
Therefore, to utilize the potential of STO as a high-$k$ capacitor or photocatalyst, it is essential to grow stoichiometric STO epitaxial films without mid-gap states. 
Since the growth of stoichiometric STO entails fine-tuning of multiple growth conditions, including the supplied flux ratio of Ti and Sr, the growth temperature, and oxidation strength in the case of molecular beam epitaxy (MBE), only a few papers have reported highly insulating stoichiometric STO films having the same lattice constants as bulk STO and no absorption in the band gap.\cite{28lee2016,29flores2017} 

The conventional trial-and-error approach to optimizing the growth conditions is time-consuming and costly, and the reproducibility of optimization depends on the individual researcher. 
In contrast, data-driven decision-making approaches have attained high-throughput in experiments where machine learning models, such as Bayesian optimization (BO) and artificial neural networks, are incrementally updated by newly measured data.\cite{30mueller2015,31lookman2016,32burnaex2015,33agrawal2016,34ueno2018,35ren2018,36wakabayashi2018,37li2018,38hou2019,39xue2016,add_baird2022} 
BO is a sample-efficient approach for global optimization,\cite{40Snoek} 
which has proven itself useful for streamlining the optimization of the thin film growth conditions.\cite{41wakabayashi2019,42takiguchi2020,43shimizu2020,44wakabayashi2022} 
However, a technical challenge for growth optimization has remained. Namely, the search procedure needs to find a suitable parameter reliably since experiments are costly in terms of time, labor, and expense.
To this end, exploration needs to be encouraged, especially when the suitable parameter region lies in a complex shape in the search space. 
Such a complex-shaped growth parameter space may force the search method to become stuck at suboptimal growth condition parameters.

In this study, to obtain highly insulating stoichiometric STO films, 
we develop a new BO method that encourages the exploration by adapting the hyperparameter of the prediction model to get out of suboptimal parameters. 
We demonstrate the efficacy of our adaptation first by using simulated data and then through implementation for real materials growth: 
machine-learning-assisted molecular beam epitaxy (ML-MBE) of STO films. Results of MBE growth and crystallographic analyses of grown samples are accumulated to produce the next growth conditions with BO (Fig.~\ref{fig:MLMBE_flow}).
As a result, we developed highly insulating stoichiometric STO films with lattice constants identical to that of bulk STO. 
Visible-to-UV light spectroscopy shows no optical absorption in the band gap, and the films were achieved in only $44$ MBE growth runs. 
The reproducible highly insulating stoichiometric STO films will contribute to the development of the next generation of electronic and photonic devices. 

\section{\label{sec:method}Methods}
\subsection{\label{sec:BO}Bayesian optimization with adaptive prior mean}
This section outlines how BO tackles the optimization problem and its adaptation of the prior mean function. 
Detailed formulations are presented in Appendix~\ref{supp:bo}. 
BO is a method for optimizing a black box function in the form of $y=f({\bf x})$, 
where function $f:\mathbb{R}^D \rightarrow \mathbb{R}$ is unknown and expensive to evaluate given $D$-dimensional input ${\bf x}\in \mathpzc{X} \subset \mathbb{R}^D$ of a specified search space $\mathpzc{X}$. 
In materials growth optimization, ${\bf x}$ and $y$ represent the growth parameters and physical properties used to evaluate grown materials, respectively. 
Examples of physical properties include electrical resistance and X-ray diffraction intensity. 
BO searches the parameter space by repeating the following steps: (i) Construct a prediction model based on the Gaussian process (GP)\cite{rasmussen2006} given the past $n$ observations $\mathpzc{D}_n=\left\{({\bf x}_i, y_i) \right\}_{i=1}^{n}$, 
(ii) evaluate an acquisition function to find a promising ${\bf x}'$ that is likely to give a good function output, and 
(iii) evaluate the new point ${\bf x}_{n+1}={\bf x}'$ and acquire its function value $y_{n+1}$ to update the prediction model using $\mathpzc{D}_{n+1}=\mathpzc{D}_n \cup \left\{ (y_{n+1},{\bf x}_{n+1}) \right\}$. 

The GP uses a prior mean function $\eta({\bf x})$ for predicting the value of $f({\bf x}')$ at unseen ${\bf x}'$. 
Nevertheless, this function is typically fixed as $\eta({\bf x})=0$ or a certain constant~\cite{shahriari2016,frazier2018,40Snoek} throughout the BO iterations because of the black box nature of $f$.
Our method still employs a constant function in the form of $\eta({\bf x})=m_0$,  
but it adaptively calculates a hyperparameter $m_0$ using past observations $\mathpzc{D}_n$. 
Figure~\ref{fig:m0_difference} visualizes an example of difference in prediction of outcome at unseen parameters caused by different choices of $m_0$ (see also Appendix \ref{supp:eta}). 
This example uses the two-dimensional Ackley function as $f({\bf x})$ [Fig.~\ref{fig:m0_difference}~(a)], where the search progress has focused on the left-top region. 
Since this function has four isolated peaks with different heights, 
BO needs to explore the search space instead of persisting in one of the peaks. 
In the example in Fig.~\ref{fig:m0_difference}, the $m_0=0$ case keeps on searching the skirt of left-top sub-optimal peak [Fig.~\ref{fig:m0_difference}~(d)]. 
In contrast, the $m_0 \approx 0.2$ case jumps into the unexplored left-bottom area [Fig.~\ref{fig:m0_difference}~(e)]. 
While greater $m_0$ encourages the exploration, fixing $m_0$ at a large value may not always be efficient: 
a large $m_0$ tends to produce an optimistic prediction in an unexplored parameter space. 
This can trigger unnecessary explorations leading to a plateau of the sub-optimal function value. 
Thus, the choice of $m_0$ needs to take account of the balance between the exploration and exploitation in the parameter search. 

To mitigate this dilemma, we developed three methods called adaptive leveling (AL), empirical Bayes (EB) and empirical Bayes uniform (EBu). 
The AL method draws $m_0$ from the uniform distribution between the minimum and maximum of the past observations to obtain a balance between the exploitation and exploration by the stochastic choice. 
We also developed the EB and EBu methods to justify the modification of hyperparameter of prior using the observations. 
Their further descriptions are deferred to Appendix~\ref{supp:al}. 

\subsection{\label{sec:ML-MBE}ML-MBE growth and sample characterizations}
\begin{figure*}[t]
\centering
\includegraphics[width=0.8\linewidth]{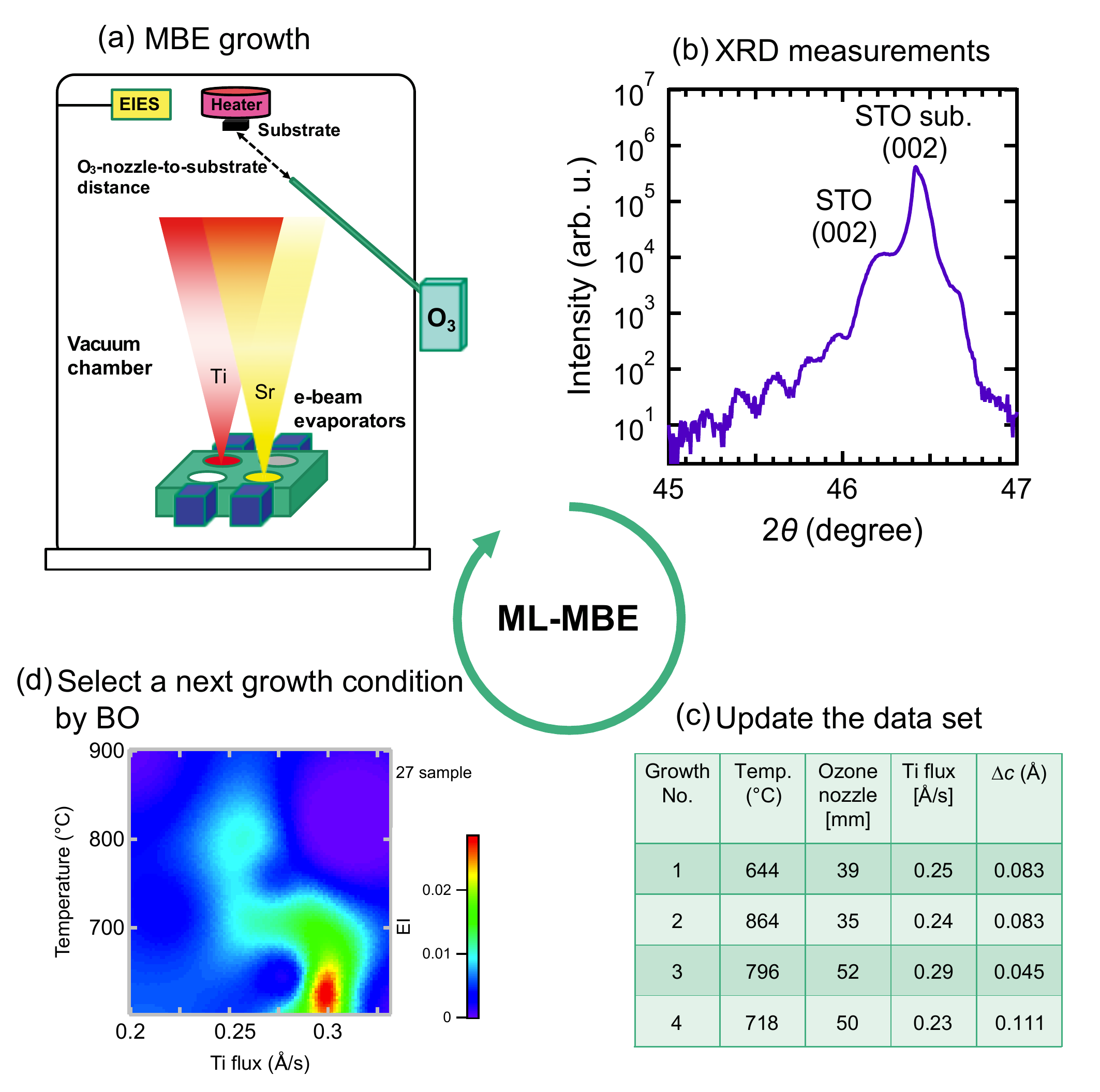}
\caption{\label{fig:MLMBE_flow}
Flow of ML-MBE growth using BO. 
(a) Schematic illustration of our multisource oxide MBE setup. EIES: Electron Impact Emission Spectroscopy. 
(b) X-ray diffraction (XRD) $\theta$--$2\theta$ scan of the STO film with a $\Delta c$ of 0.045~\AA, as an example. STO sub. means the peak from the substrate.
(c) Growth conditions for four samples, as an example. 
(d) Two-dimensional plots of EI values at the O$_3$-nozzle-to-substrate distance of 65.5 mm obtained from the collected data for 27 samples, as an example.
}
\end{figure*}

Epitaxial STO films with a thickness of 60 nm were grown on STO (001) substrates in a custom-designed MBE system with multiple e-beam evaporators [Fig. \ref{fig:MLMBE_flow}~(a)]. 
Detailed information about the MBE system is described elsewhere.\cite{45naito1995,46yamamoto2013,47wakabayashi2022,48wakabayashi2019,add_wakabayashi2021} 
We precisely controlled the Sr and Ti elemental fluxes by monitoring the flux rates with an electron-impact-emission-spectroscopy sensor and feeding the results back to the power supplies for the e-beam evaporators. 
The oxidation during growth was carried out with ozone (O$_3$) gas ($\sim15\%$ O$_3$ $+ 85\%$ O$_2$) introduced through an alumina nozzle pointed at the substrate. 
For the stoichiometric STO growth, it is important to fine tune the growth conditions (the ratio of the Ti flux to the Sr flux, growth temperature, and local ozone pressure at the growth surface).\cite{27kumar2020,49naito1998,50jalan2009,51ohnishi2005}
To systematically change the Ti flux ratio to the Sr flux, 
we changed the Ti flux while keeping the Sr flux at 0.98~\AA/s.
The growth temperature was controlled by the heater shown in Fig.~\ref{fig:MLMBE_flow}~(a). 
We can adjust the local ozone pressure at the growth surface by changing the O$_3$-nozzle-to-substrate distance [Fig.~\ref{fig:MLMBE_flow}~(a)] 
while keeping the flow rate of O$_3$ gas at $\sim0.25$ sccm. 

We executed the BO algorithm in a three-dimensional space. 
The search windows for the Ti flux rate, growth temperature, and O$_3$-nozzle-to-substrate distance were 
0.20--0.33 \AA/s, 600--900$^\circ$C, and 10--80 mm, respectively. 
We searched equally spaced grid points for each parameter. 
The number of points 
of the respective quantities was 100. 
Since the three-dimensional parameter space consisted of 1,000,000 
($100^3$) points, 
performing a trial for the entire space in a point-by-point manner is unrealistic, 
as only several runs can be carried out per day with a typical MBE system. 
In order to evaluate the stoichiometry of the films, we measured $\theta$--$2\theta$ scanned x-ray diffraction (XRD) [Fig.~\ref{fig:MLMBE_flow}~(b)] 
since the increase in the lattice constant is a good indicator of the magnitude of the off-stoichiometry of STO by changes in the cation and/or oxygen concentration.\cite{49naito1998,50jalan2009,52ohnishi2008,53lebeau2009} 
Therefore, we adopted the difference in the $c$-axis lattice constant of the film and the substrate ($\Delta c$) as the evaluation value. Thermal conductivity might be considered as a useful metric for evaluating the crystalline quality of SrTiO$_3$ films with high-quality films of sufficient thickness reproducing the thermal conductivity observed in bulk single crystals.\cite{Oh2011,Brooks2015} However, to make the thermal conductivity of a SrTiO$_3$ film distinguishable from that of the SrTiO$_3$ substrate, a film thickness of several hundred nm is required,\cite{Oh2011,Brooks2015} and therefore thermal conductivity is not suitable for the evaluation of the SrTiO$_3$ films with a thickness of 60 nm used in this study. If films are thick enough and the thermal conductivity measurements are reliable, optimization using BO methods with thermal conductivity as the evaluation value should also be possible. When XRD diffractions from the STO phase were indiscernible and/or diffractions from SrO, TiO$_2$, or Sr$_{n+1}$Ti$_n$O$_{3n+1}$ ($n$: integer; $n$$\neq$$\infty$) Ruddlesden-Popper series\cite{54lee2013} precipitates (impurity phases) appeared, 
we defined the evaluation value of those samples to be the worst experimental $\Delta c$ value by that time. 
This imputation of the missing data generated when the designated phase is not formed enabled a direct search of the wide three-dimensional parameter space.\cite{44wakabayashi2022} 

Here, a black box function $\Delta c=f({\bf x})$ is the target function specific to our STO films, 
and ${\bf x}$ represents the growth parameters (Ti flux rate, growth temperature, and O$_3$-nozzle-to-substrate distance). 
We used data $\mathpzc{D}_n={({\bf x}_i,y_i) }_{i=1}^n$ [Fig.~\ref{fig:MLMBE_flow}~(c)] obtained from past $n$ MBE growths and XRD measurements [Fig.~\ref{fig:MLMBE_flow}~(b)] of STO films to construct a model to predict the value of $f({\bf x})$ at an unseen ${\bf x}$. 
To this end, we used the GP to estimate the mean $m$ and variance $s^2$ at an arbitrary parameter value ${\bf x}$ (see Appendix~\ref{supp:gp} for details). 
Specifically, the GP predicts the value of $f({\bf x})$ as a Gaussian-distributed variable $\mathpzc{N}(m({\bf x}), s^2({\bf x}))$, where $m$ and $s^2$ depend on ${\bf x}$ and $\mathpzc{D}_n$. 
In short, $m({\bf x})$ and $s^2({\bf x})$ represent the expected value and uncertainty of $\Delta c$ at ${\bf x}$. 
To consider the inherent noise in the $\Delta c$ of STO films grown under nominally the same conditions, 
the variance of the observation noise $\sigma_{\varepsilon}^2$ of the GP model was set to $0.01$. 
In our implementation, we used the Mat\'{e}rn $\frac{5}{2}$ kernel since it is good at fitting functions with steep gradients.\cite{40Snoek} 
We iterated the routine after the initial MBE growth with five random initial growth parameters and XRD measurements. 
First, the GP was updated using the data set at the time [Fig.~\ref{fig:MLMBE_flow}~(c)]. 
Subsequently, to assign the value of the growth parameter in the next run, we calculated the expected improvement (EI) [Fig.~\ref{fig:MLMBE_flow}~(d)].\cite{55mockus1978}

\section{\label{sec:result}Results and Discussion}
\subsection{\label{sec:sim-exp}Experiments with simulated data}
\begin{figure*}[p!]
\centering
\includegraphics[width=0.58\linewidth]{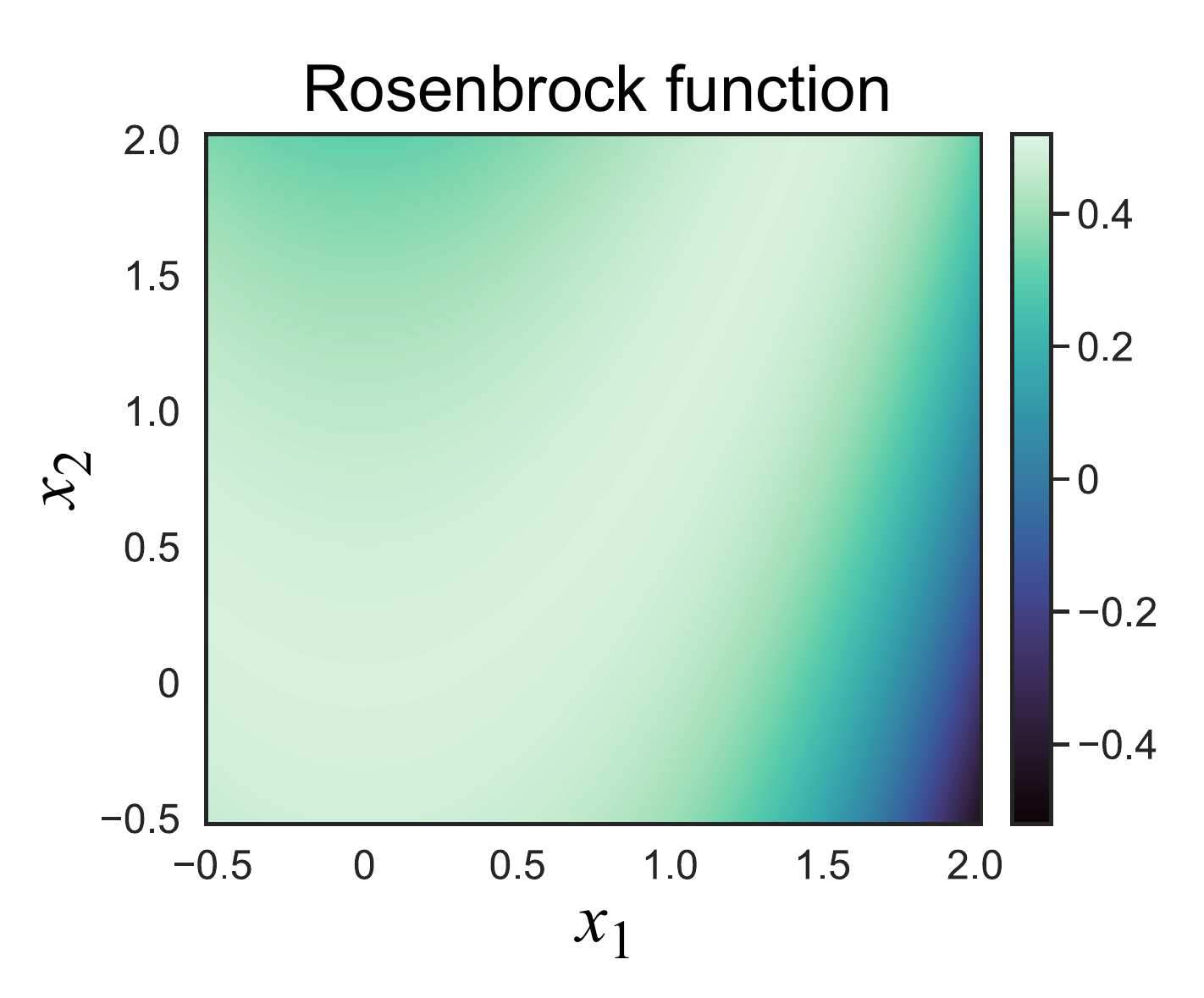}
\caption{\label{fig:objective-rb}
Rosenbrock function with $D=2$ employed for simulated optimization experiment. } 
\end{figure*}

This section investigates the optimization performance for simulated functions by five methods: 
the baseline with $m_0=0$, a simple adaptation of $m_0$ by taking the average of observed data $m_0 = \frac{1}{n}\sum_{i=1}^n y_i$ referred to as {\em DA}, which stands for data averaging, and the methods that adjust $m_0$: AL, EB, and EBu. 
We use two functions—the Ackley function\cite{ackley1987,molga2005} and the Rosenbrock function\cite{rosenbrock1960,molga2005}. 
The boundary of search space $\mathpzc{X}$ was set at $-0.5$ and $2$ for each element. 
These functions allow us to set an arbitrary number of dimensions $D$. 
In this study, we used $D=2, 4,$ and 6. 
Generally speaking, larger dimensionality $D$ makes black box optimization more challenging. 
The Ackley function with $D=2$ is illustrated in Figure~\ref{fig:m0_difference}~(a) and the Rosenbrock function with $D=2$ is displayed in Figure~\ref{fig:objective-rb}, respectively. 
For each configuration, 
all methods (baseline, DA, AL, EB, and EBu) were iterated until 100 observations were obtained. 
These optimization processes were repeated five times with five randomly chosen initial observations. 
Each evaluation contained noise with $\sigma_{\varepsilon}^{2}=0.001$. 
These functions are described in Appendix~\ref{supp:obj}. 

Figure~\ref{fig:exp-simulated} shows the optimization results for the Ackley and Rosenbrock functions  with $D=2, 4$, and 6. 
Each curve indicates the best observation value $\displaystyle \left( \max_{1\leq i \leq n}y_i \right)$ averaged the over five runs as a function of the number of observations $n$. 
The shaded area indicates the best and worst observations among the five runs. 
A curve that rises with fewer observations indicates a better search algorithm, one that requires fewer resources before reaching a high value of $y$. 
A vertically narrow shaded area means that the method performs robustly against randomness in the search process, 
such as the initial choice of parameters and random seeds. 

For the Ackley function with $D=2$ [Fig.~\ref{fig:exp-simulated}~(a)], all methods reached the top peak on average. 
In particular, the baseline, DA and AL found the optimal value after 50 observations in all five trials of the optimization process. 
Since EB and EBu were stuck at the second-best peak occasionally, their average performance was slightly inferior to that of the baseline, DA and AL. 
In larger dimensionalities $D=4$ and $6$ for the Ackley function [Figs.~\ref{fig:exp-simulated}~(b) and \ref{fig:exp-simulated}~(c)], the baseline method struggled to improve. 
This is because the baseline tends to be trapped at one of the suboptimal peaks. 
In contrast, DA, AL, EB, and EBu with adaptive $m_0$ clearly outperformed the baseline. 
This supports the efficacy of the variable prior mean for BO. 
Nevertheless, none of the methods attained the maximum value of 0.5. 
This explains the general difficulty of black box optimization in high-dimensional search space, 
especially when the objective function has multiple and disconnected peaks. 
The performance of DA was worse in $D=6$ [Fig.~\ref{fig:exp-simulated}~(c)]. 
Since DA method tends to stabilize the value of $m_0$ with more observations, the algorithm failed to explore novel regions. 

Figures~\ref{fig:exp-simulated}~(d)--\ref{fig:exp-simulated}~(f) show the optimization results for the Rosenbrock function with $D=2, 4$, and 6. 
Unlike the case with the Ackley function, all methods performed comparably well under all conditions. 
These results mean that the optimization of the Rosenbrock function is easy due to its concave surface [Fig.~\ref{fig:objective-rb}]. 
Since the configuration of search space $\mathpzc{X}$ was restricted to a limited region between -0.5 and 2, a high evaluation value $f({\bf x}) \geq 0.3$ was present for a large portion of the search space. 
This allowed all methods to work reasonably well in our experiment. 
With that said, DA method showed a slower improvements when $D=4$ [Fig.~\ref{fig:exp-simulated}~(e)] and $D=6$ [Fig.~\ref{fig:exp-simulated}~(f)]. 
This is because an occasional observation of low $y_i$ decreases $m_0$ of DA, which caused conservative and pessimistic prediction that led to slower improvements. 

Among AL, EB and EBu that change $m_0$ at each step, the performance was similar in most configurations. 
With that said, some trials of EB and EBu were inferior to that of AL in, for example, the optimization result of the Ackley functions with $D=2$ and $6$ and the Rosenbrock function with $D=4$. 
Owing to the stability and reliability of the performance, we adopted the AL method 
for the ML-MBE growth of STO films.

\subsection{\label{sec:sto-exp}Application to ML-MBE of STO film}
\begin{figure*}[p]
\centering
\includegraphics[width=0.95\linewidth]{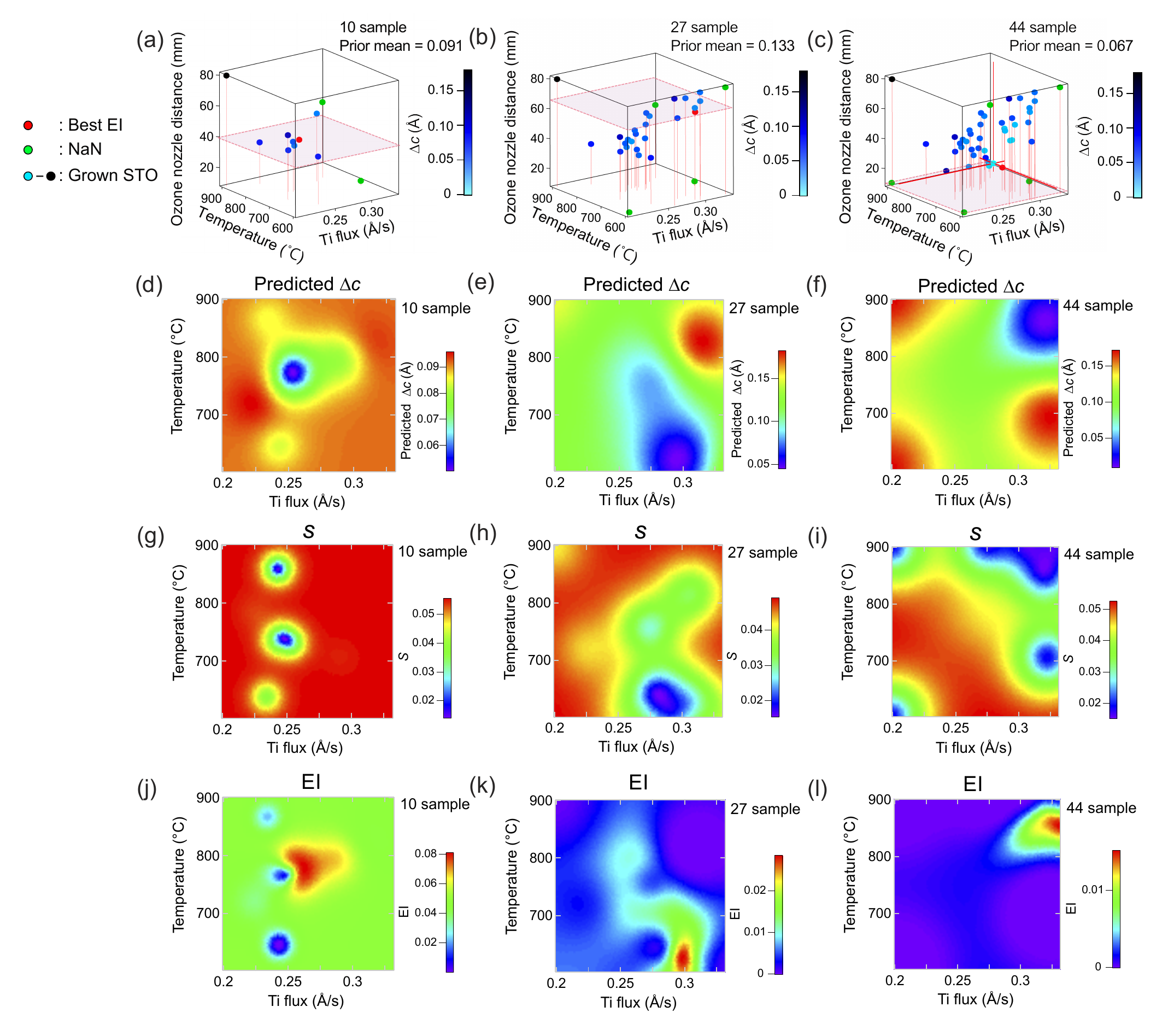}
\caption{\label{fig:MLMBE_results}
(a)--(c) Experimental $\Delta c$ values in the three-dimensional growth parameter space for 10 (a), 27 (b), and 44 (c) samples. 
The green spheres indicate the \texttt{NaN} points at which the STO phase was not obtained. 
The red spheres indicate the most promising conditions with the highest EI values, which should be examined in the next growth run. 
The red planes indicate the cutting plane of the O$_3$-nozzle-to-substrate distance, at which the highest EI value was obtained. 
(d)--(l): Two-dimensional plots of predicted $\Delta c$ values (d)--(f), $s$ values (g)--(i), and EI values (j)--(l) at O$_3$-nozzle-to-substrate distance of 39.5~mm [(d),(g),(j)], 65.5~mm [(e),(h),(k), and~10 mm [(f),(i),(l)], 
which were obtained from the collected data for 10 [(d),(g),(j)], 27 [(e),(h),(k)], and 44 [(f),(i),(l)] observations, respectively. 
The O$_3$-nozzle-to-substrate distance was that at which the highest EI value was obtained.
}
\end{figure*}

\begin{figure*}[p]
\centering
\includegraphics[width=0.5\linewidth]{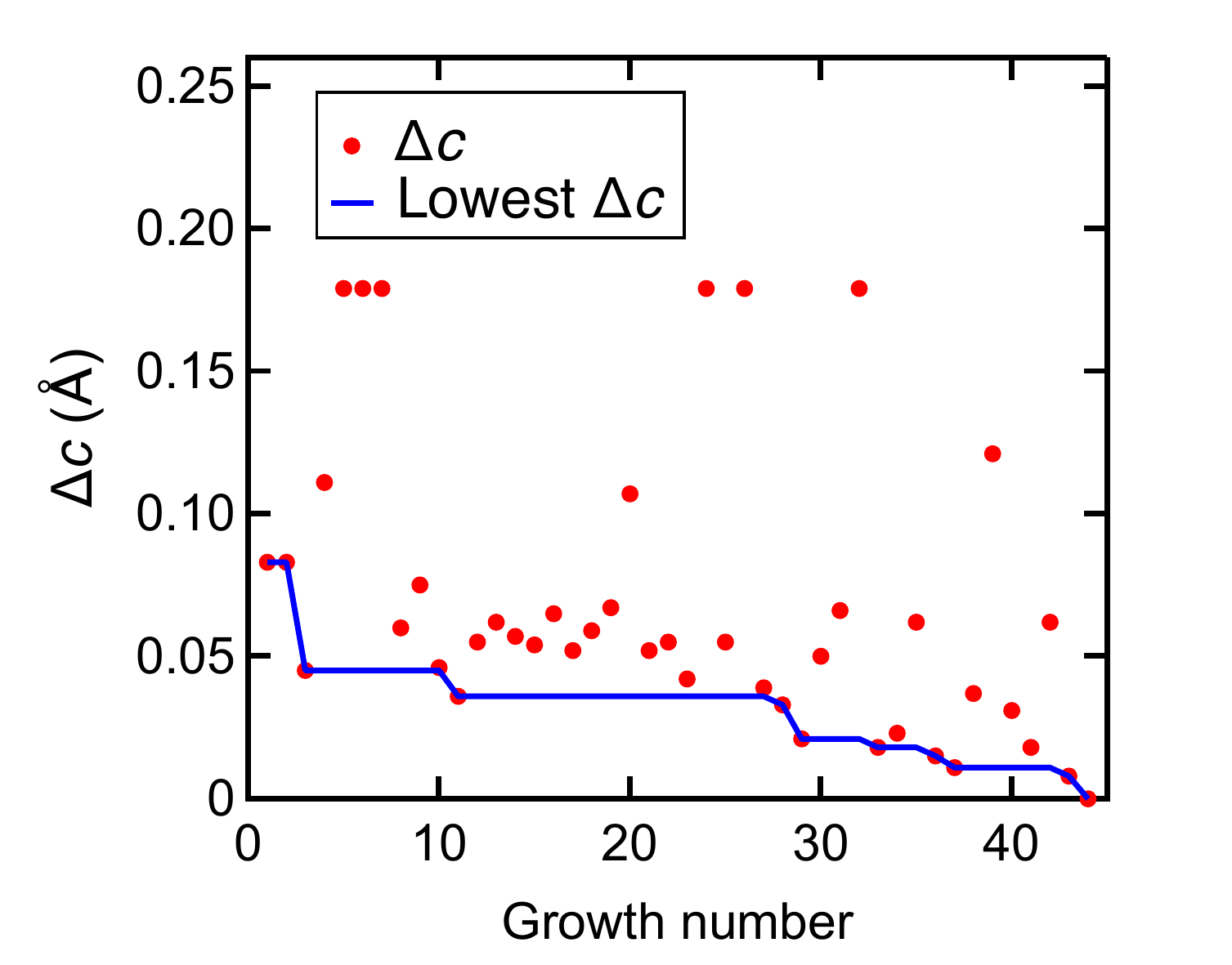}
\caption{\label{fig:MLMBE_run_vs_c}
Actual $\Delta c$ values and lowest experimental $\Delta c$ plotted as a function of growth number. 
}
\end{figure*}

To obtain stoichiometric STO films with no absorption in the band gap, we grew STO films by the AL method implemented in ML-MBE. 
Figure~\ref{fig:MLMBE_results} shows how the BO algorithm predicts $\Delta c$ values with unseen parameter configurations and acquires new data points. 
The process starts with five random initial growth parameters and gains experimental $\Delta c$ values for the updated GP model with 10 [Fig.~\ref{fig:MLMBE_results}~(a)], 27 [Fig.~\ref{fig:MLMBE_results}~(b)] and 44 [Fig.~\ref{fig:MLMBE_results}(c)] observations. 
Two-dimensional plots of the predicted $\Delta c$, $s({\bf x})$ and EI values at the O$_3$-nozzle-to-substrate distance, at which the highest EI value was obtained, are shown in the lower panels [Figs.~\ref{fig:MLMBE_results}~(d)--\ref{fig:MLMBE_results}~(l)]. 
Within the first ten samples from the start, the STO phase had not formed at two growth conditions [green spheres in Fig.~\ref{fig:MLMBE_results}~(a)]. 
Thus, we defined the $\Delta c$ value of these samples as the worst experimental one at that time. 
This imputation of experimental failure enabled a direct search of the wide three-dimensional parameter space.\cite{44wakabayashi2022} 
According to the GPR prediction from the ten samples, 
the highest EI was obtained at the Ti flux $= 0.26$~\AA/s, growth temperature $= 777~^\circ$C, and O$_3$-nozzle-to-substrate distance $= 39.5$ mm [Fig.~\ref{fig:MLMBE_results}~(a)]. 
This predicted growth condition yielded the $\Delta c$ of $0.036$~\AA, smaller than the minimum value of $0.045$~\AA~at that time [Fig.~\ref{fig:MLMBE_results}~(b)]. The more surrounding data points there are, the smaller $s$ becomes, and the less there are, the larger it becomes. Therefore, the region with relatively small $s$ of the predicted $\Delta c$ became larger as the number of experimental samples increased from 10 to 44 [Figs.~\ref{fig:MLMBE_results}~(g)--~\ref{fig:MLMBE_results}~(i)], meaning that the prediction accuracy had increased, resulting in the lower EI values [Figs.~\ref{fig:MLMBE_results}~(j)--\ref{fig:MLMBE_results}~(l)]. 
Through this optimization process, in which the exploration is encouraged by varying the prior mean, 
the lowest $\Delta c$ value decreased and reached an ideal value of 0 in only 44 MBE growth runs (Fig.~\ref{fig:MLMBE_run_vs_c}). 
The ideal $\Delta c$ of 0 was achieved at the Ti flux $= 0.32$~\AA/s, growth temperature $= 852~^\circ$C, and O$_3$-nozzle-to-substrate distance $= 13.5$ mm [Fig.~\ref{fig:MLMBE_results}~(c)]. 
The achievement of the target material with the desired properties in such a small number of optimizations demonstrates the efficacy of the AL method for high-throughput materials growth. 

The optimized Ti flux ($0.32$~\AA/s) corresponds to the Ti/Sr ratio of 1.026, which is near the stoichiometric value of 1. This may be due to the low vapor pressure of the most stable strontium oxide SrO and titanium oxide TiO$_2$,\cite{Guguschev2014} for which the sticking coefficients of both Sr and Ti become almost 1—even if SrO and TiO$_2$ are concomitantly formed during the growth of SrTiO$_3$, they do not desorb from the growth surface and are eventually transformed to SrTiO$_3$. Generally, in complex oxide films, if the sticking coefficients of each constituent cation were known as functions of growth temperature and oxidation strength, the optimum supplied flux ratio could be predicted, at least in principle. However, one cannot optimize the whole growth conditions by predicting or just using reasoning from a crystal growth perspective. Instead, the combined impact of the growth temperature, the oxidation strength, and the supplied flux ratio on the oxide film growth can only be obtained empirically through experiments. This is because actual crystal growth dynamics are complicated and cannot be comprehended even when thermodynamic phase diagrams are available. The proposed method enables high-quality and efficient materials growth, independent of the researcher's knowledge and experience, even in cases where such prior knowledge is limited.
\subsection{\label{sec:STO-analysis}Crystallographic and optical properties of stoichiometric STO films} 
\begin{figure*}[p]
\centering
\includegraphics[width=0.95\linewidth]{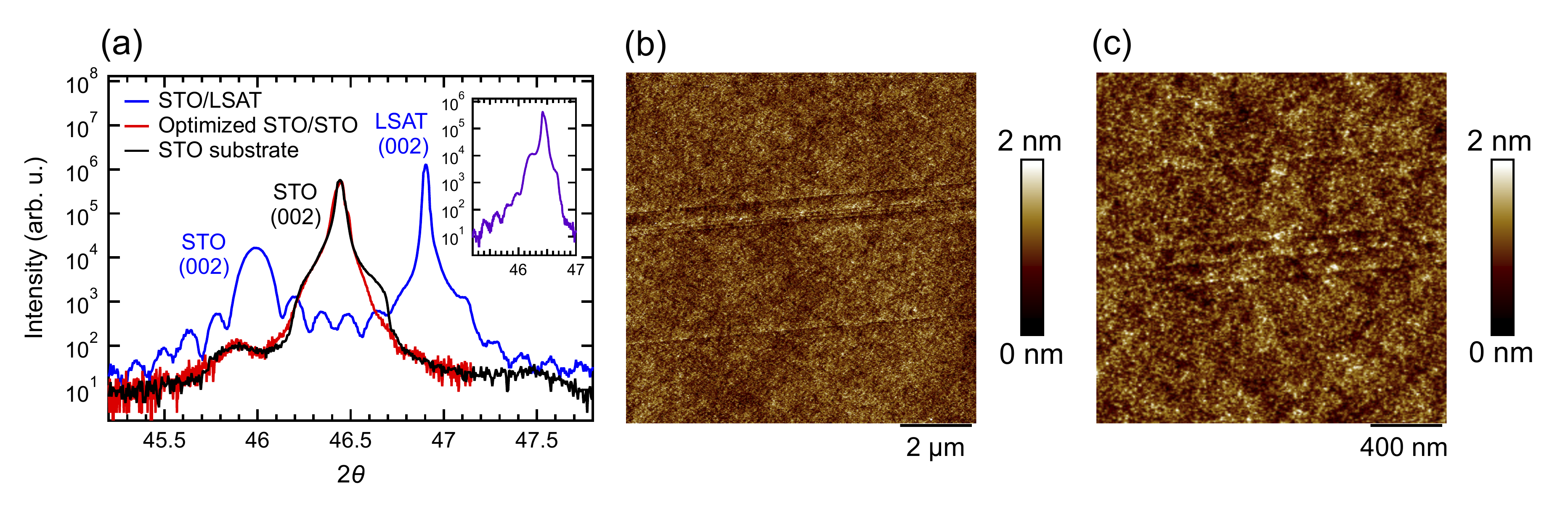}
\caption{\label{fig:XRD_AFM}
(a) XRD $\theta$--$2\theta$ scans of the stoichiometric STO film grown on (001) STO and LSAT substrates. Inset shows the XRD $\theta$--$2\theta$ scan of the off-stoichiometric STO film.
(b) AFM image of the stoichiometric STO film on (001) STO. 
(c) Magnified image of (b).
}
\end{figure*}

We experimentally characterized the physical properties of the stoichiometric STO film with $\Delta c$ of 0. 
For comparison, we also examined the physical properties of the off-stoichiometric STO film with $\Delta c$ of 0.045~\AA~  
grown under one of the first random growth conditions (Ti flux $= 0.29$~\AA/s, growth temperature $= 796~^\circ$C, and O$_3$-nozzle-to-substrate distance $= 52$~mm). 
The sheet resistance was measured by a standard two-point method with Ag electrodes deposited on the STO surface. 
The stoichiometric STO is highly insulating, exceeding the measurable range (sheet resistance $> 50$~M$\Omega$), while the off-stoichiometric STO film shows a relatively small sheet resistance of $1.4$~k$\Omega$. 
This result indicates that precise stoichiometry adjustment is necessary to obtain highly resistive STO. 
The crystallinity of the STO films was examined by XRD, atomic force microscopy (AFM), 
and scanning transmission electron microscopy (STEM). 
Figure~\ref{fig:XRD_AFM}~(a) shows XRD $\theta$--$2\theta$ scans of the stoichiometric STO film around the (002) STO Bragg peak. 
The XRD $\theta$--$2\theta$ scans of the STO film grown on (001) (LaAlO$_3$)$_{0.3}$(SrAl$_{0.5}$Ta$_{0.5}$O$_3$)$_{0.7}$ (LSAT) under the same growth conditions are also shown. 
It has been reported that non-stoichiometric STO films have larger lattice constants than stoichiometric ones.~\cite{27kumar2020,49naito1998,50jalan2009,51ohnishi2005,52ohnishi2008,56brooks2009} 
The XRD pattern for the stoichiometric STO on STO shows a good overlap between film and substrate peaks without XRD fringes. 
The lack of fringes in the XRD data—which would be observed for finite repetition of the unit cell—is a typical feature of stoichiometric STO films~\cite{27kumar2020,52ohnishi2008} since the films merge with the substrate and become indistinguishable. In contrast, the fringes are clearly observed for the films heteroepitaxially grown on the LSAT substrate, which allows for thickness estimation of the SRO films. The film thickness estimated from the periods of the Laue fringes (62 nm) agrees very well with that calculated by the Sr flux rate (60 nm), whose sticking coefficient is 1. 
Figures~\ref{fig:XRD_AFM}~(b) and \ref{fig:XRD_AFM}~(c) show the AFM images of the stoichiometric STO film. 
The root-mean-square roughness is 0.25 nm, indicating that the stoichiometric STO film has smooth surfaces. 

\begin{figure*}[p]
\centering
\includegraphics[width=0.95\linewidth]{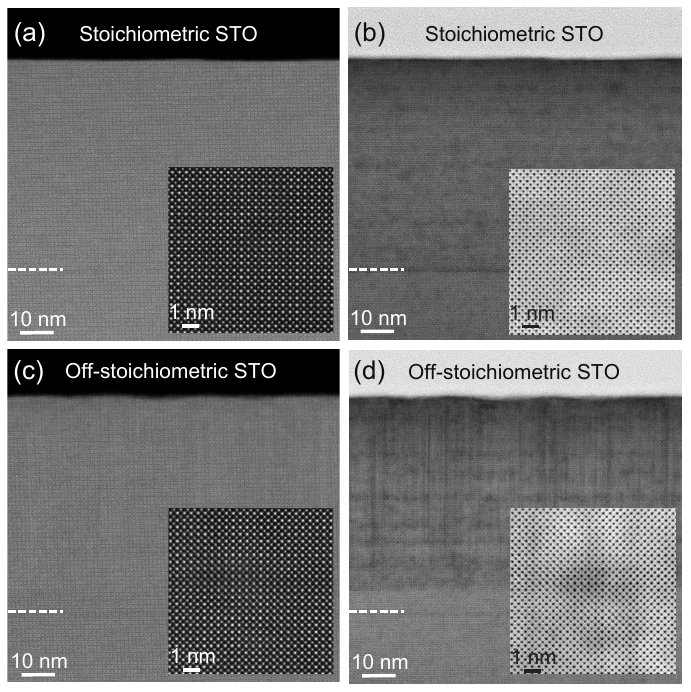}
\caption{\label{fig:STEM}
(a),~(c) HAADF-STEM and (b),~(d) ABF-STEM images of the stoichiometric [(a),~(b)] and off-stoichiometric [(c),~(d)] STO films along the [100] direction. 
Dashed lines indicate the interfaces between the grown STO layers and the substrates.
Insets show magnified images at the center of the films.
}
\end{figure*}

Figure~\ref{fig:STEM} shows high-angle annular dark-field (HAADF)- and annular bright-field (ABF)-STEM images of the stoichiometric and off-stoichiometric STO films taken with a JEOL JEM-ARM 200F microscope. 
Since the intensity in the HAADF-STEM image is proportional to $\sim\mathrm{Z}^n$ ($n\sim$  1.7--2.0, and Z is the atomic number),\cite{57pennycook1991} 
the brighter spheres and darker ones in Figs.~\ref{fig:STEM}~(a) and \ref{fig:STEM}~(c)] are assigned to Sr- ($\mathrm{Z} = 38$) and Ti- ($\mathrm{Z} = 22$) occupied columns, respectively. The ABF-STEM images [\ref{fig:STEM}~(b) and \ref{fig:STEM}~(d)] represent atomic arrangement of oxygen since oxygen is emphasized in annular bright-field ABF-STEM images.\cite{Okunishi2009} The film and the substrate are nearly indistinguishable in the HAADF-STEM image for the stoichiometric film [Fig.~\ref{fig:STEM}~(a)], 
indicating the ideal cationic arrangement at the interface. 
In contrast, the threading dislocations perpendicular to the film surface are observed in the ABF-STEM image for the off-stoichiometric film [Fig.~\ref{fig:STEM}~(d)], which are merely observed in the HAADF-STEM image as well [Fig.~\ref{fig:STEM}~(c)]. 
Such threading dislocations have been reported in Sr-rich STO films and are thought to be Ruddlesden–Popper planar faults.\cite{56brooks2009} 
In addition, the magnified ABF-STEM image [inset in Fig.~\ref{fig:STEM}~(d)] reveals strong contrast due to local atomic dechanneling.\cite{58lee2016,59li2021} 
Since oxygen is emphasized in ABF-STEM unlike HAADF-STEM images [Fig.~\ref{fig:STEM}~(c)], 
the contrasts in the ABF-STEM image should come from the oxygen vacancies. 
The oxygen vacancies in the off-stoichiometric STO film are consistent with the growth conditions with an oxidation strength lower than that for the stoichiometric STO film (O$_3$-nozzle-to-substrate distances are $13.5$ and $52$ mm for the stoichiometric STO and off-stoichiometric STO films, respectively). 

\begin{figure*}[p]
\centering
\includegraphics[width=0.55\linewidth]{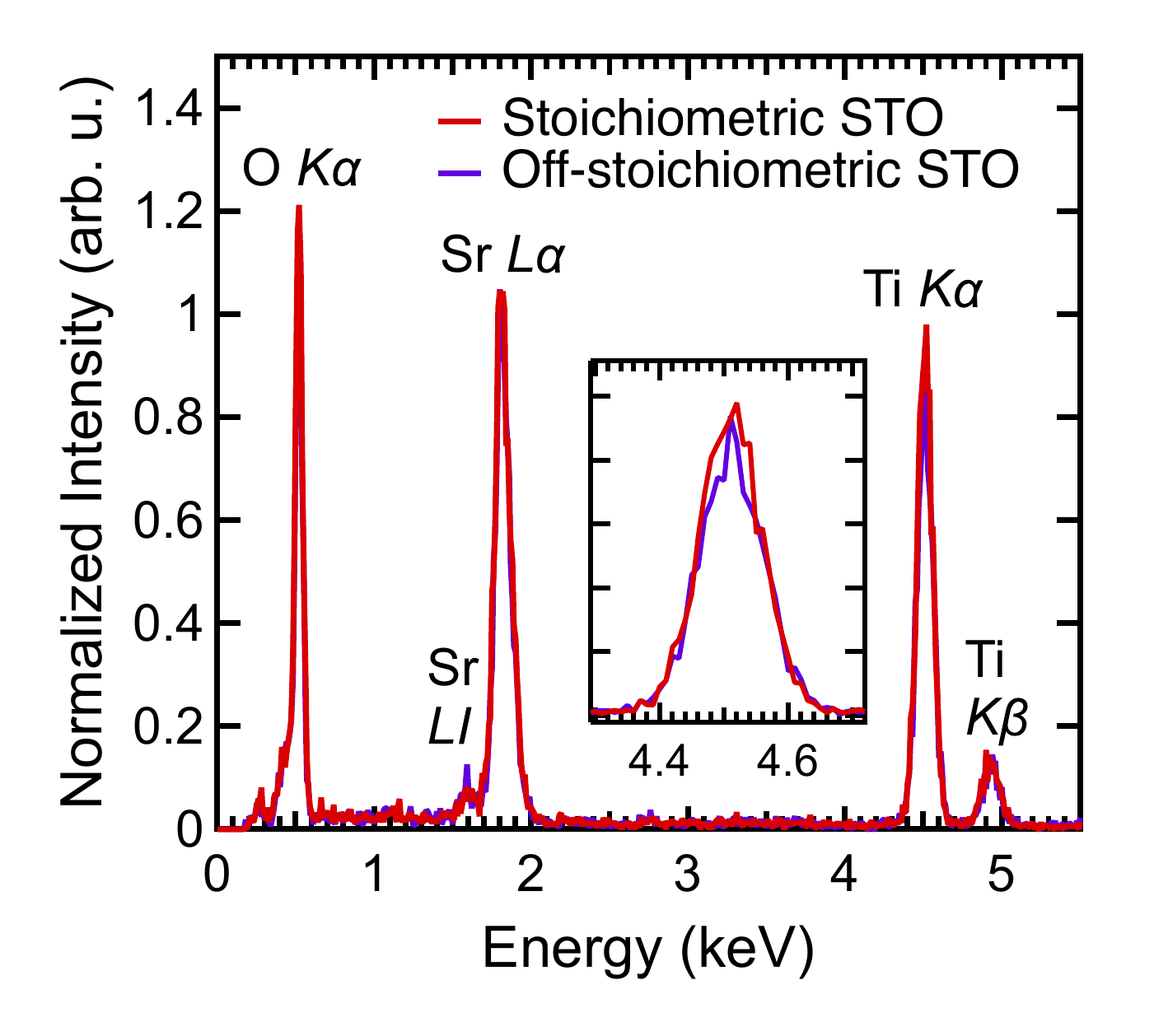}
\caption{\label{fig:EDS}
EDS spectra for the stoichiometric and off-stoichiometric STO films. 
The inset shows a magnified view at the Ti $K\alpha$. 
The spectra are normalized at the Sr $L\alpha$ peak intensities for easy comparison.
}
\end{figure*}

To determine the Ti/Sr composition ratios, we carried out energy dispersive x-ray spectroscopy (EDS) measurements taken also with a JEOL JEM-ARM 200F microscope. 
Figure~\ref{fig:EDS} shows the EDS spectra for the stoichiometric and off-stoichiometric STO films. 
Only Sr, Ti, and O peaks are observed in both films, indicating no observable impurities in the films. 
The Ti/Sr composition ratio was estimated by the Ti $K\alpha$/Sr $L\alpha$ integrated intensity ratios normalized by that of the STO substrate, {\em i.e.}, 
the Ti/Sr ratio of the STO substrate is assumed to be 1. 
The estimated Ti/Sr ratios in the stoichiometric and off-stoichiometric films are 1.00 and 0.94, respectively. 
Note that typical accuracy of the EDS for the Sr and Ti $L\alpha$ integrated intensity is $\pm$0.01--$\pm$0.04.\cite{60gao2018,61alonso2019}

\begin{figure*}[p!]
\centering
\includegraphics[width=0.55\linewidth]{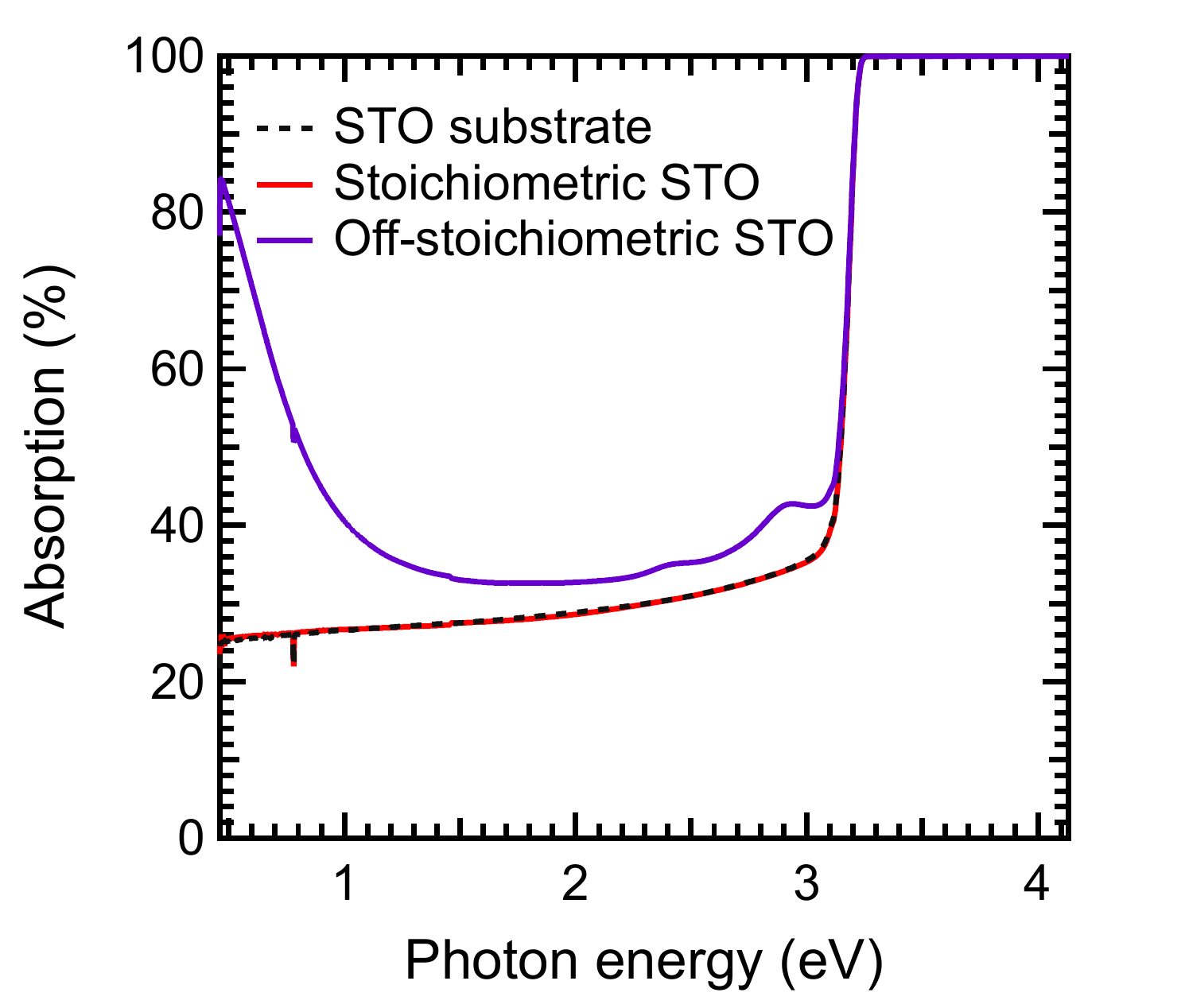}
\caption{\label{fig:absorption}
Optical absorptions of the STO substrate and the stoichiometric and non-stoichiometric STO films on STO at room temperature. 
}
\end{figure*}

Figure~\ref{fig:absorption} shows the optical absorption of the stoichiometric and off-stoichiometric STO films at room temperature. 
The sudden increase of the absorption at 3.2~eV originates from the O 2$p$-to-Ti 3$d$ charge transfer transition in STO films and substrates, 
indicating the bandgap of STO.\cite{07konta2004}
The absorption spectrum of the stoichiometric STO is identical to that of the STO substrate,
indicating that the stoichiometric STO could be an ideal mother material for photocatalysis applications. 
In contrast, non-stoichiometric STO shows the Drude (free-electron-carrier) absorption (photon energy $<1$ eV) and absorption from deep-impurity states ($1$~eV $<$ photon energy $<3.2$~eV).\cite{27kumar2020} 
The absorptions at 2.4 and 2.9~eV may originate from the excitation of electrons trapped by oxygen vacancies since it is widely observed in reduced STO.\cite{57pennycook1991}

\section{\label{sec:conclusion}Conclusion}
We demonstrated the stoichiometric growth of STO films via Bayesian optimization with an adaptive hyperparameter of a prior mean function. 
To obtain highly insulating stoichiometric STO films, 
we developed a new BO method that encourages the exploration by adjusting the prior mean to get out of suboptimal parameters. 
Using simulated data, we found the efficacy of all the methods that vary the prior mean value, reproducibly reaching the global best conditions. 
Among these methods, we employed the AL method for ML-MBE. 
In only 44 MBE growth runs, our approach attained highly insulating stoichiometric STO films having no absorption in the band gap, which will contribute to the next generation of electronic and photonic devices. 
The proposed algorithm provides an efficient experimental design platform that is not as dependent on the experience and skills of individual researchers. 
It will enhance the efficiency of not only oxide electronics but also various material syntheses and autonomous syntheses.\cite{burger2020,ziatdinov2022,stach2021} 


\newpage
\section*{Authors' Contributions}
{\bf Y. K. Wakabayashi}: Conceptualization (equal); Methodology (equal); Software (supporting); Validation (equal); Investigation (lead); Supervision (equal); Writing -- Original Draft (equal); Writing -- Review \& editing (equal). 
{\bf T. Otsuka}: Conceptualization (equal); Methodology (equal); Software (lead); Validation (equal); Investigation (supporting); Supervision (equal); Writing -- Original Draft (equal); Writing -- Review \& editing (equal). 
{\bf Y. Krockenberger}: Investigation (supporting); Writing -- Review \& editing (supporting). 
{\bf H. Sawada, Y. Taniyasu, H. Yamamoto}: Writing -- Review \& editing (supporting).

\section*{Data Availability}

The data and code that support the findings of this study are openly available in GitHub at \url{https://github.com/nttcslab/adaptive-leveling-BO}.



\begin{figure*}[p]
\centering
\includegraphics[width=0.98\linewidth]{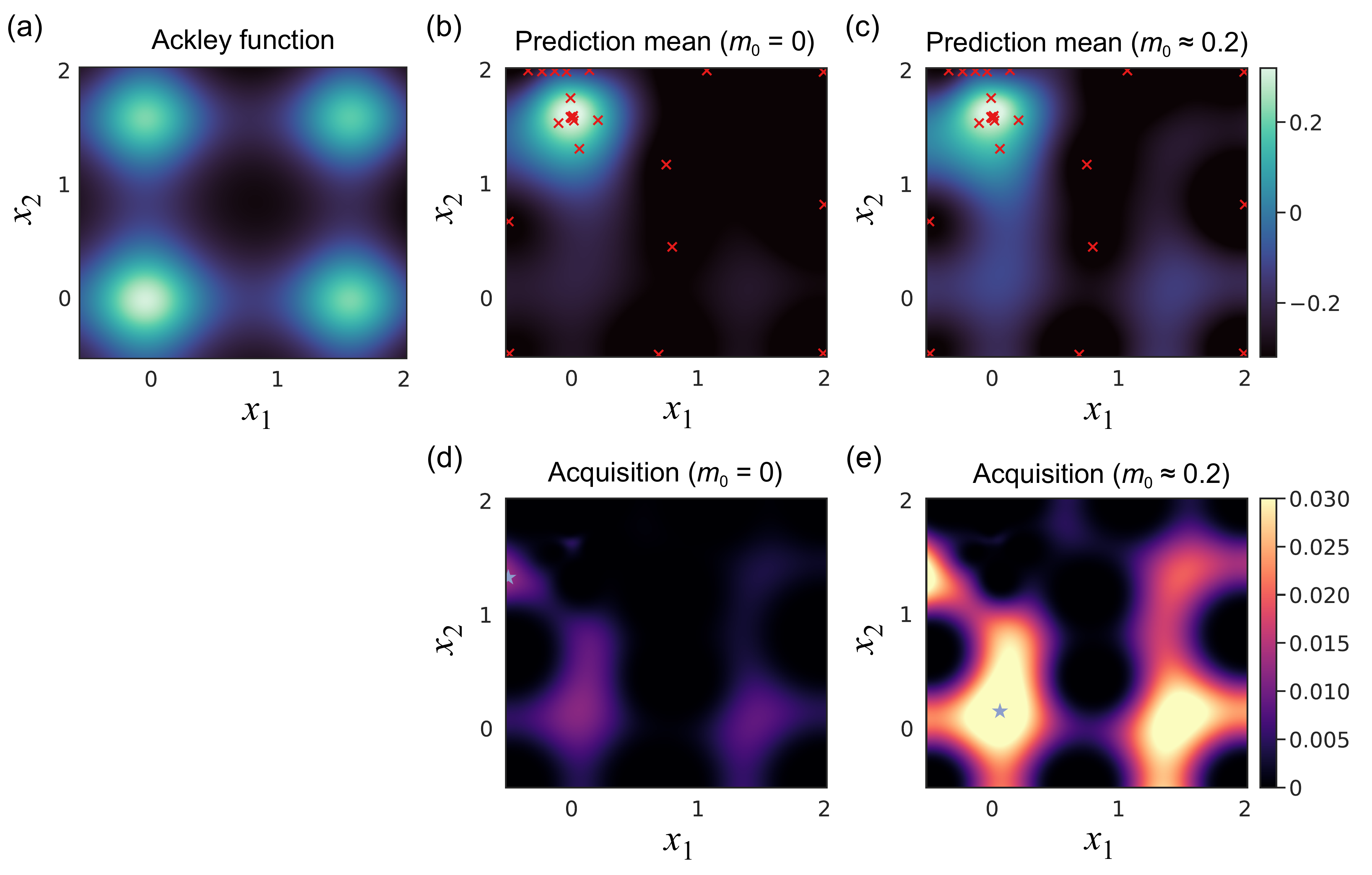}
\caption{\label{fig:m0_difference}
(a) Ackley function on two-dimensional parameter space [$x$$_1$,$x$$_2$]. Among four peaks, the left-bottom one at $[0,0]^\top$ gives the largest value. 
(b), (c) Predicted mean $m({\bf x})$ and (d), (e) acquisition function $a_\mathrm{EI}({\bf x})$ with 20 observations depicted with $\times$ marks. 
In (d) and (e), the $\star$ mark represents the position of maximum of $a_\mathrm{EI}({\bf x})$. 
(b), (d) Results with $m_0 = 0$. (c), (e) Results with $m_0 \approx 0.2$.
}
\end{figure*}



\newpage





\begin{figure*}[p]
\centering
\includegraphics[width=0.95\linewidth]{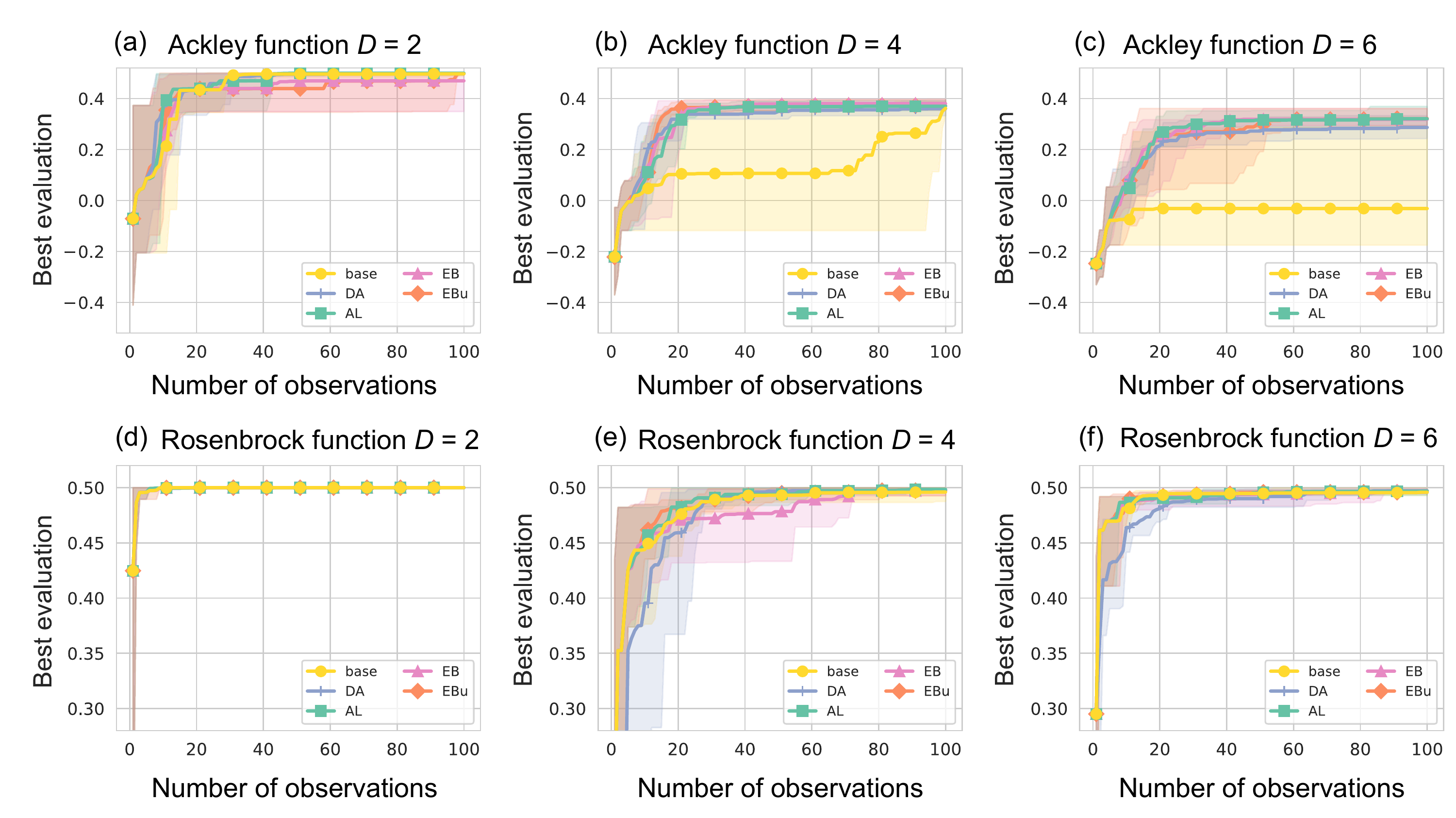}
\caption{\label{fig:exp-simulated}
(a)--(c) Optimization results for the Ackley function with (a) $D=2$, (b) $D=4$, and (c) $D=6$. 
(d)--(f) Optimization results for the Rosenbrock function with (d) $D=2$, (e) $D=4$, and (f) $D=6$. 
Solid lines indicate the best evaluation value found so far as a function of number of observations averaged over five trials. 
Shaded areas means the best and worst performance among the five trials. 
}
\end{figure*}

\clearpage

\appendix
\section{\label{supp:bo}Bayesian optimization with adaptive prior mean function}
Bayesian optimization (BO) is a method for optimizing a black box function $f({\bf x})$:
\begin{align}
    \max_{{\bf x}\in \mathpzc{X}} f({\bf x}), \label{eq:maxf}
\end{align}
where we can evaluate the function value given a $D$-dimensional parameter ${\bf x}\in \mathpzc{X} \subset \mathbb{R}^D$ as: 
\begin{align}
    y &= f({\bf x}) + \varepsilon, \label{eq:observe}
\end{align}
but the underlying function $f({\bf x})$ is unknown. 
In materials growth  optimization, ${\bf x}$ and $y$ represent growth parameters and physical properties used to evaluate grown materials, respectively. 
Examples of physical properties include electrical resistance and X-ray diffraction intensity. 
Here, we assume observation y contains an additive Gaussian noise $\varepsilon\sim\mathpzc{N}(0, \sigma_{\varepsilon}^2)$. 
While we will present the maximization problem, note that we can equivalently cope with the minimization problem by negating the observation. 
Furthermore, we assume the search space $\mathpzc{X}$ is bounded so that each element in ${\bf x}$ may be normalized between 0 and 1, as described in existing work.\cite{44wakabayashi2022} 
In addition, we may combine the floor padding trick\cite{44wakabayashi2022} to handle experimental failures where the observation of Eq.~\eqref{eq:observe} is unavailable. 
BO optimizes the black box function by constructing a prediction model to find an unexplored parameter with a good chance to improve its function value. 
Then, the predicted model is iteratively updated after obtaining the actual observation corresponding to the parameter. 
This process is typically iterated until a certain function value is attained or the number of observations reaches a certain budget for the experiment. 

\subsection{\label{supp:gp}GP prediction and acquisition function}
The Gaussian process is adopted for the prediction model. 
This model predicts the outcome of unseen parameters as a normal distribution. 
\begin{align}
    p(y'|{\bf x}', \mathpzc{D}_n) &= \mathpzc{N}\left( m({\bf x}'), s^2({\bf x}') \right), \label{eq:prediction}
\end{align}
where the past $n$ observations are denoted as $\mathpzc{D}_n=\left\{({\bf x}_1, y_1), \dots, ({\bf x}_n, y_n) \right\}$. 
The predictive mean $m({\bf x}')$ and variance $s^2({\bf x}')$ are calculated using the kernel $k(\cdot, \cdot)$ as follows:
\begin{align}
    m({\bf x}') &= \eta({\bf x}') + {\bf k}_n^\top {\bf K}_n^{-1} \left( {\bf y}_n - \pmb{\eta}_n \right), \label{eq:pred-m} \\ 
    s^2({\bf x}') &= k({\bf x}', {\bf x}') - {\bf k}_n^\top {\bf K}_n^{-1} {\bf k}_n, \label{eq:pred-s}
\end{align}
where the $i$th element of vector ${\bf k}_n \in \mathbb{R}^n$ is given by $k({\bf x}_i, {\bf x}')$ and 
the element at the $i$th row and $j$th column of Gram matrix ${\bf K}_n\in\mathbb{R}^{n \times n}$ is $k({\bf x}_i, {\bf x}_j)$. 
We used the Mat\'{e}rn $\frac{5}{2}$ kernel for our GP. 
Vector ${\bf y}_n$ represents the past observations; 
${\bf y}_n = \left[y_1, \dots, y_n \right]^\top$ with $\cdot^\top$ means the transpose operator. 
Function $\eta({\bf x})$ is the prior mean function that shifts the predictive mean by subtracting $\pmb{\eta}_n= \left[\eta({\bf x}_1), \dots, \eta({\bf x}_n) \right]^\top$ from observation ${\bf y}_n$ 
and adding $\eta({\bf x}')$ back to predictive mean $m({\bf x}')$
Intuitively, when parameter ${\bf x}'$ is distant from any of observed data ${\bf x}_1, \dots, {\bf x}_n$, 
the predictive mean is dominated by the prior mean function, {\em i.e.}, $m({\bf x}') \approx \eta({\bf x}')$ 
because elements of ${\bf k}_n$ approach zero with distance-based kernels, such as the Mat\'{e}rn or radial basis function kernels. 

Given the predictions, the acquisition function is evaluated to decide which parameter to employ in the next trial. 
We use the expected improvement (EI) criterion 
\begin{align}
    a_\mathrm{EI} &= \mathbb{E}_{p\left(y' | {\bf x}', \mathpzc{D}_n \right)} \left[ \left( y' - \bar{y}_n \right) \mathbb{I}_{y' \geq \bar{y}_n} \right], \label{eq:ei}
\end{align}
where $\displaystyle \bar{y}_n = \max_{1\leq i \leq n}y_i$ is the largest observation and 
$\mathbb{I}_{y' \geq \bar{y}_n}$ is the indicator function that equals 1 when $y' \geq \bar{y}_n$ and 0 otherwise. 
This acquisition function evaluates the expectation of improvement over the best observation $\bar{y}_n$ at unseen parameter ${\bf x}'$. 

\subsection{\label{supp:eta}Impact of prior mean function on optimization}
The choice of $\eta({\bf x})$ will affect the acquisition function as well as the parameter search efficiency. 
The typical choice of the prior mean function is $\eta({\bf x})=0$, 
particularly because we lack the knowledge of underlying {\em black box} function $f({\bf x})$. 
We replace the prior mean with a constant function in the form of $\eta({\bf x})=m_0$ since a more flexible functional form of $\eta$ to potentially approximate $f$ is inaccessible.


Figure~\ref{fig:m0_difference} shows the difference in the prediction mean and the acquisition for the two-dimensional Ackley function with $m_0=0$ and $m_0 \approx 0.2$. 
See Appendix~\ref{supp:obj} and Eq.~\eqref{eq:ackley} for the description of the Ackley function. 
As shown in Fig.~\ref{fig:m0_difference}~(a), the Ackley function $f({\bf x})$ has four peaks with the left-bottom one at ${\bf x}=[0,0]^\top$ being the highest. 
In this example, the left-top suboptimal peak has been intensively searched [Figs.~\ref{fig:m0_difference}~(b) and \ref{fig:m0_difference}~(c)]. 
Comparison with $m_0=0$ [Fig.~\ref{fig:m0_difference}~(b)] and $m_0\approx 0.2$ [Fig.~\ref{fig:m0_difference}~(c)] shows that the predictive mean $m({\bf x}')$ with $m_0\approx 0.2$ is larger than that with $m_0=0$ in the half bottom region. 
This difference in the predicted mean results in the difference in the acquisition function [Figs.~\ref{fig:m0_difference}~(d) and \ref{fig:m0_difference}~(e)].
The use of $m_0>0$ increases the predictive mean of unexplored region and leads to finding other peaks. 

While a greater $m_0$ worked favorably in the example in Fig.~\ref{fig:m0_difference}, large $m_0$ may not always be efficient: 
a large $m_0$ tends to produce an optimistic prediction in unexplored regions. 
This can trigger unnecessary explorations leading to a plateau at the suboptimal function value. 
Thus, the choice of hyperparameter $m_0$ needs to take account of the balance between the exploration and exploitation in the parameter search. 

\subsection{\label{supp:al}Methods for adapting prior mean function hyperparameter}
The discussion in Section~\ref{supp:eta} motivates us to adapt $m_0$. 
A simple way is to average observed data $m_0 = \frac{1}{n}\sum_{i=1}^n y_i$. 
We call this approach DA (data averaging). 
Despite its simplicity, DA works well in practice [Fig.~\ref{fig:exp-simulated}], but tends to saturate $m_0$ to a certain value as more observations are accumulated. 
This can compensate the exploration of parameter search. 
In the following, we present three methods with varying $m_0$ for efficient BO. 

The first method is adaptive leveling (AL) that randomly sets $m_0$ between the best and worst observations every time a new observation is acquired: 
\begin{align}
    m_0 \sim \mathpzc{U}\left(\min_{1\leq i \leq n} y_i, \max_{1\leq i \leq n} y_i \right), \label{eq:al}
\end{align}
where $\mathpzc{U}(l, h)$ denotes the uniform distribution on interval $[l, h]$. 
We expect that $m_0$ is not always too optimistic or pessimistic by the stochastic choice. 

While the design of AL is intuitive, this may be unnatural in light of Bayesian inference: $m_0$, the hyperparameter of the prior mean, is specified after observing data. 
The empirical Bayes method\cite{casella1985,carlin2000} can justify the choice of $m_0$ based on observations ${\bf y}_n$ by maximizing the marginal likelihood. 
The marginal log likelihood, or the evidence, of the Gaussian process for data $\mathpzc{D}_n$ is given as follows:\cite{rasmussen2006} 
\begin{align}
    \ln p({\bf y}_n | {\bf x}_1, \dots, {\bf x}_n) &= -\frac{1}{2} \left({\bf y}_n - {\bf 1}_n m_0 \right)^\top {\bf K}_n^{-1} \left({\bf y}_n - {\bf 1}_n m_0 \right) + \mathrm{const}. \label{eq:marginal}
\end{align}
Here, ${\bf 1}_n$ is an $n$-dimensional vector with all elements being 1 and the constant term in Eq.~\eqref{eq:marginal} includes quantities irrelevant to $m_0$. 
The marginal likelihood is maximized when 
\begin{align}
    m_0 &= \frac{{\bf 1}_n^\top  {\bf K}_n^{-1} {\bf y}_n}{{\bf 1}_n^\top  {\bf K}_n^{-1} {\bf 1}_n}. \label{eq:eb}
\end{align}
This quantity can be interpreted as a weighted average of elements in ${\bf y}_n$, where the weight is given by ${\bf K}_n^{-1} {\bf 1}_n$. 
We refer to this choice of $m_0$ as EB that stands for empirical Bayes.
We may optionally add some randomness to $m_0$ by considering the condition when the marginal log likelihood improves over the baseline configuration $m_0=0$. 
The empirical Bayes uniform (EBu) draws $m_0$ as follows: 
\begin{align}
    m_0 & \sim \left\{ 
    \begin{array}{lr}
        \mathpzc{U}\left(0, 2 \frac{{\bf 1}_n^\top  {\bf K}_n^{-1} {\bf y}_n}{{\bf 1}_n^\top  {\bf K}_n^{-1} {\bf 1}_n} \right) & \mathrm{if}\;\; \frac{{\bf 1}_n^\top  {\bf K}_n^{-1} {\bf y}_n}{{\bf 1}_n^\top  {\bf K}_n^{-1} {\bf 1}_n} \geq 0, \\
        \mathpzc{U}\left(2 \frac{{\bf 1}_n^\top  {\bf K}_n^{-1} {\bf y}_n}{{\bf 1}_n^\top  {\bf K}_n^{-1} {\bf 1}_n}, 0 \right) & \mathrm{otherwise}.
    \end{array}
    \label{eq:ebu}
    \right.
\end{align}
EBu has a mechanism similar to AL for switching between the exploration and exploitation, as well as a guarantee for a better marginal likelihood compared with the baseline $m_0=0$. 
In contrast, AL may use a value of $m_0$ with less marginal likelihood; its range is restricted to the range of past observations. 
This can circumvent an extreme choice of $m_0$ that may cause unstable search. 

Indeed, EB and EBu may choose $m_0 \in \left[\min_{1\leq i \leq n}y_i, \max_{1\leq i \leq n}y_i  \right]$. 
This is because some elements in ${\bf K}_n^{-1}{\bf 1}_n$ can be negative, and thus their weighted average can produce the extrapolation of observations in ${\bf y}_n$. 
Since EBu uses an extended interval between 0 and $\displaystyle 2 \frac{{\bf 1}_n^\top  {\bf K}_n^{-1} {\bf y}_n}{{\bf 1}_n^\top  {\bf K}_n^{-1} {\bf 1}_n}$, 
$m_0$ of EBu has more chance to be outside of the past observations.

\section{\label{supp:obj}Objective functions used in simulated data experiments}
The Ackley function is defined as follows:
\begin{align}
    f_{\mathrm A}({\bf x}) &\propto C_1 \left(\exp\left(-\sqrt{\frac{\sum_{d=1}^D x_d^2}{25 D}} \right) + \exp\left(\cos \left(4 \frac{\sum_{d=1}^D x_d}{D} \right) \right)  \right) + C_2, \label{eq:ackley}
\end{align}
where $x_d$ denotes the $d$th element of ${\bf x}$ and $C_1$ and $C_2$ are set such that $f_{\mathrm A}({\bf x})$ ranges from -0.5 to 0.5 on $\mathpzc{X}$. 
This function has $2^D$ peaks in $\mathpzc{X}$ and takes its maximum value at ${\bf x}={\bf 0}$. 

The Rosenbrock function is defined as follows: 
\begin{align}
    f_{\mathrm R} & \propto -C_3 \left(\sum_{d=1}^D (x_d - 1)^2 + 5 \sum_{d=1}^{D-1} \left(  x_d - x_{d+1}^2 \right)^2  \right) + C_4, \label{eq:rosenbrock}
\end{align}
where $C_3$ and $C_4$ are configured such that $f_{\mathrm R}({\bf x})$ ranges from -0.5 to 0.5 on $\mathpzc{X}$. 
This function has a valley-shaped surface with its unique peak located at ${\bf x}={\bf 1}$. 

\section{\label{supp:lattice}Optimization of deviation in lattice constant}
In the STO film growth experiment, our objective was to minimize $\Delta c$, the absolute difference of lattice constants. 
While $\Delta c \geq 0$ by definition, the prediction of GP assumes $\Delta c$ may take a negative value by fitting the normal distribution. 
We may consider two types of fix for this mismatch. 
First one is to handle the logarithm of difference: $y=-\log \Delta c$. 
Note that we negated the difference since BO is presented as a maximization of $y$. 
While this naturally maps the nonnegative value to a real value, a large difference $\Delta c$ is distorted and less emphasized. 
The second approach is to truncate the probability of $\Delta c < 0$, that is, $p(y>0|{\bf x},\mathpzc{D}_n)=0$, where $y=-\Delta c$. 
We can modify the EI criterion to derive an analytic form of 
\begin{align}
    a_\mathrm{EI}({\bf x}) &= \mathbb{E}_{\tilde{p}\left(y' | {\bf x}', \mathpzc{D}_n \right)} \left[ \left( y' - \bar{y}_n\right) \mathbb{I}_{y' \geq \bar{y}_n}  \right], \label{eq:ei-trunc}
\end{align}
where the modified predictive probability is 
\begin{align}
    \tilde{p}(y | {\bf x}, \mathpzc{D}_n) & \propto \left\{
    \begin{array}{cc}
        p(y |  {\bf x}, \mathpzc{D}_n) & \mathrm{if}\;\; y\leq 0,  \\
        0 & \mathrm{otherwise}.
    \end{array}
    \label{eq:predict-trunc}
    \right.
\end{align}

We carried out an experiment for minimizing the difference between the function value and its target.
We compared four methods: three of them minimizes the absolute difference between the target and function value, while the other uses a special acquisition function that evaluates the closeness of the prediction to the target value. 
The first method is the most straightforward one; it simply handles the absolute difference between the target and function value: $y=-|t - f({\bf x})|$, where $t$ denotes the target value. 
We call this method {\em absolute} here. 
The second method ({\em log absolute}) maximizes $y=-\log|t - f({\bf x})|$ to let $y$ take a negative value. 
The third method ({\em truncation}) modifies the acquisition function as in Eq.~\eqref{eq:ei-trunc} using $y=-|t - f({\bf x})|$. 
The last method~\cite{urenholt19} fits the GP to the function directly using $y = f({\bf x})$ to form the predictive mean $m({\bf x}')$ and variance $s^2({\bf x}')$ as in Eq.~\eqref{eq:prediction}. 
Letting the squared distance denote $d({\bf x}')= \left(t - f({\bf x}')\right)^2$,  
we model the distance using 
the non-central chi-squared distribution $\chi_\mathrm{NC}^2$ as 
\begin{align}
    \left. \frac{d}{s^2({\bf x}')} \right| {\bf x}', \mathpzc{D}_n & \sim \chi_\mathrm{NC}^2 \left(\lambda({\bf x}') \right), \label{eq:ncchi2}
\end{align}
where the degree of freedom of $\chi_\mathrm{NC}^2$ is 1 and its non-centrality is  $\lambda({\bf x}') = \left(\frac{t - m({\bf x}')}{s({\bf x}')}\right)^2$.
The acquisition function evaluates how $d$ can be close to zero by, for example, assessing the distance at predetermined percentile of the distribution in Eq.~\eqref{eq:ncchi2}. 
We refer to the last method as {\em chi$^2$}.

\begin{figure*}[p]
\centering
\includegraphics[width=0.98\linewidth]{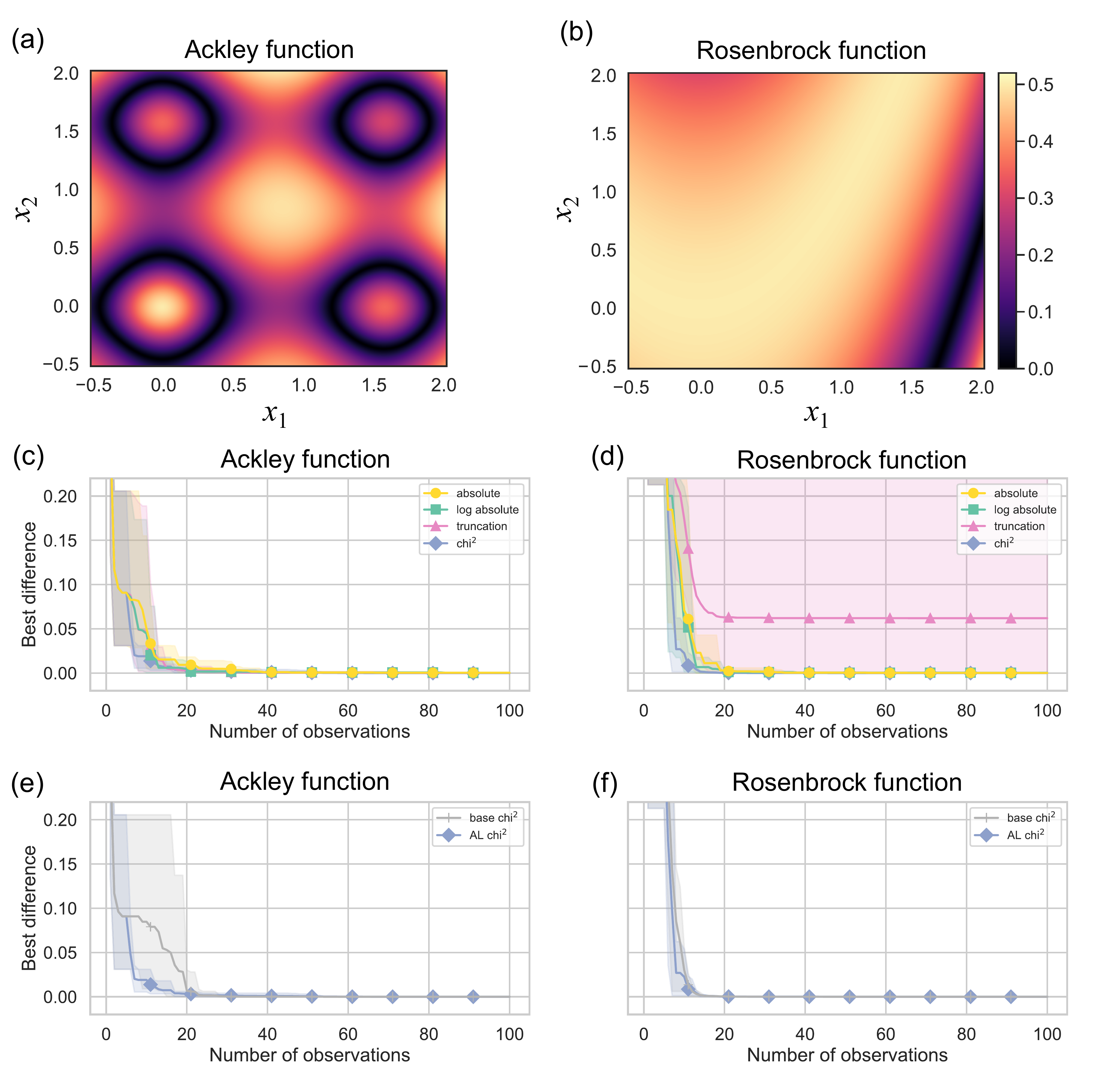}
\caption{\label{fig:abs_diff}
(a) Absolute value of Ackley function with $D=2$. The original function is displayed in Fig.~\ref{fig:m0_difference}~(a). 
(b) Absolute value of Rosenbrock function with $D=2$. The original function is displayed in Fig.~\ref{fig:objective-rb}.
(c) Optimization results of minimizing the difference between the function value and $0$, {\em i.e.}, the absolute value of the Ackley function. 
(d) Optimization results of minimizing the absolute value of the Rosenbrock function. 
(c), (d) Solid lines indicate the best evaluation value found so far as a function of the number of observations averaged over five trials. 
Shaded areas mean the best and worst performance for the five runs. 
(e) Optimization results of absolute Ackley function using $\chi_\mathrm{NC}^2$ with $m_0=0$ (base) and AL. 
(f) Optimization results of absolute Rosenbrock function using $\chi_\mathrm{NC}^2$ with $m_0=0$ (base) and AL. 
}
\end{figure*}

Our experiments used the Ackley and Rosenbrock functions with $D=2$ and set the target as $t=0$. 
Namely, we minimized the absolute value of these functions. 
Figure~\ref{fig:abs_diff} (a) and (b) show the value $|f({\bf x}) - t|$ of Ackley function and Rosenbrock function, respectively. 
Figure~\ref{fig:abs_diff} (c) and (d) show the optimization results of each function.
For both functions, the optimization process with each method was repeated five times until 100 observations acquired. 
Solid lines indicate the smallest difference between the function value and target as a function of the number of observations averaged over the five runs. 
Shaded areas mean the best and worst performance among the five runs. 
The results in Figure~\ref{fig:abs_diff} (c) and (d) were given by the AL method for GP prediction. 
In the case of the Ackley function [Fig.~\ref{fig:abs_diff}~(c)], 
all approaches eventually reached $f({\bf x})=0$ after 40 observations. 
In contrast, the truncation approach failed once out of five trials [Fig.~\ref{fig:abs_diff}~(d)]. 
The rest of the methods showed stable performance for finding $f({\bf x})=0$. 
Figure~\ref{fig:abs_diff} (e) and (f) investigate the effect of the choice of $m_0$ (AL vs. baseline) on the chi$^2$ approach. 
While AL made little difference for the Rosenbrock function [Fig.~\ref{fig:abs_diff}~(f)], 
AL accelerated the optimization in the case of the Ackley function [Fig.~\ref{fig:abs_diff}~(e)]. 
The observation in an unexplored region was encouraged when $m_0 \approx 0$ was chosen. 
This behavior enhanced the chance of finding $f({\bf x})=0$ on the wavy surface of the absolute Ackley function [Fig.~\ref{fig:abs_diff}~(a)]. 

Our results showed that all the approaches except the truncation one gave comparable performance for both functions. 
While chi$^2$ reduced the difference between the function and target values slightly quicker than the other approaches, 
we adopted the simplest absolute approach for the STO film growth experiment in expectation of its robust behavior and easier interpretation of progress in the wild environment. 
Since the MBE growth experiment requires extensive time, we leave the use of chi$^2$ for optimizing the growth parameters as future work.

\clearpage

\bibliography{STO}

\end{document}